\documentclass[letterpaper]{article}
\usepackage{isea}
\usepackage[pdftex]{graphicx}
\usepackage{times}
\usepackage{helvet}
\usepackage{courier}
\usepackage[numbers]{natbib}
\usepackage{hyperref}
\hypersetup{
  colorlinks=false,      
  pdfborder={0 0 0},     
}

\title{LAV: Audio-Driven Dynamic Visual Generation with Neural Compression and StyleGAN2}
\author{Jongmin Jung$^{1}$, Dasaem Jeong$^{2}$\\
$^{1}$Dept. of Artificial Intelligence, $^{2}$Dept. of Art \& Technology, Sogang University\\
Seoul, South Korea\\
jongmin@sogang.ac.kr, dasaemj@sogang.ac.kr\\
\newline
\newline
}
\begin{document} 
\maketitle

\begin{abstract}
This paper introduces LAV (Latent Audio-Visual), a system that integrates EnCodec’s neural audio compression with StyleGAN2’s generative capabilities to produce visually dynamic outputs driven by pre-recorded audio. Unlike previous works that rely on explicit feature mappings, LAV uses EnCodec embeddings as latent representations, directly transformed into StyleGAN2’s style latent space via randomly initialized linear mapping. This approach preserves semantic richness in the transformation, enabling nuanced and semantically coherent audio-visual translations. The framework demonstrates the potential of using pre-trained audio compression models for artistic and computational applications.
\end{abstract}

Audio-Visual Art, Neural Audio Compression, Generative Adversarial Networks, EnCodec, StyleGAN2

\section{Introduction}

Artistic and computational systems that translate audio into visual media have become increasingly sophisticated with the advent of neural networks. However, many current approaches are based on explicit feature extraction, which can lead to a loss of semantic richness in the translation process. This paper presents LAV (Latent Audio-Visual), a system that uses EnCodec's \cite{encodec} compact and semantically rich latent representations as the foundation for audio-to-visual synthesis. By directly transforming these embeddings into StyleGAN2’s \cite{Karras2019stylegan2} style latent space, LAV achieves nuanced outputs while preserving the semantic coherence of audio inputs.\footnote{LAV demo videos:\newline
\url{https://www.youtube.com/playlist?list=PL3cldJxIjpAzkgpTWlLUxAElsT-5hCHh9}}

\begin{figure}
    \includegraphics[width=3.31in, height=6in]{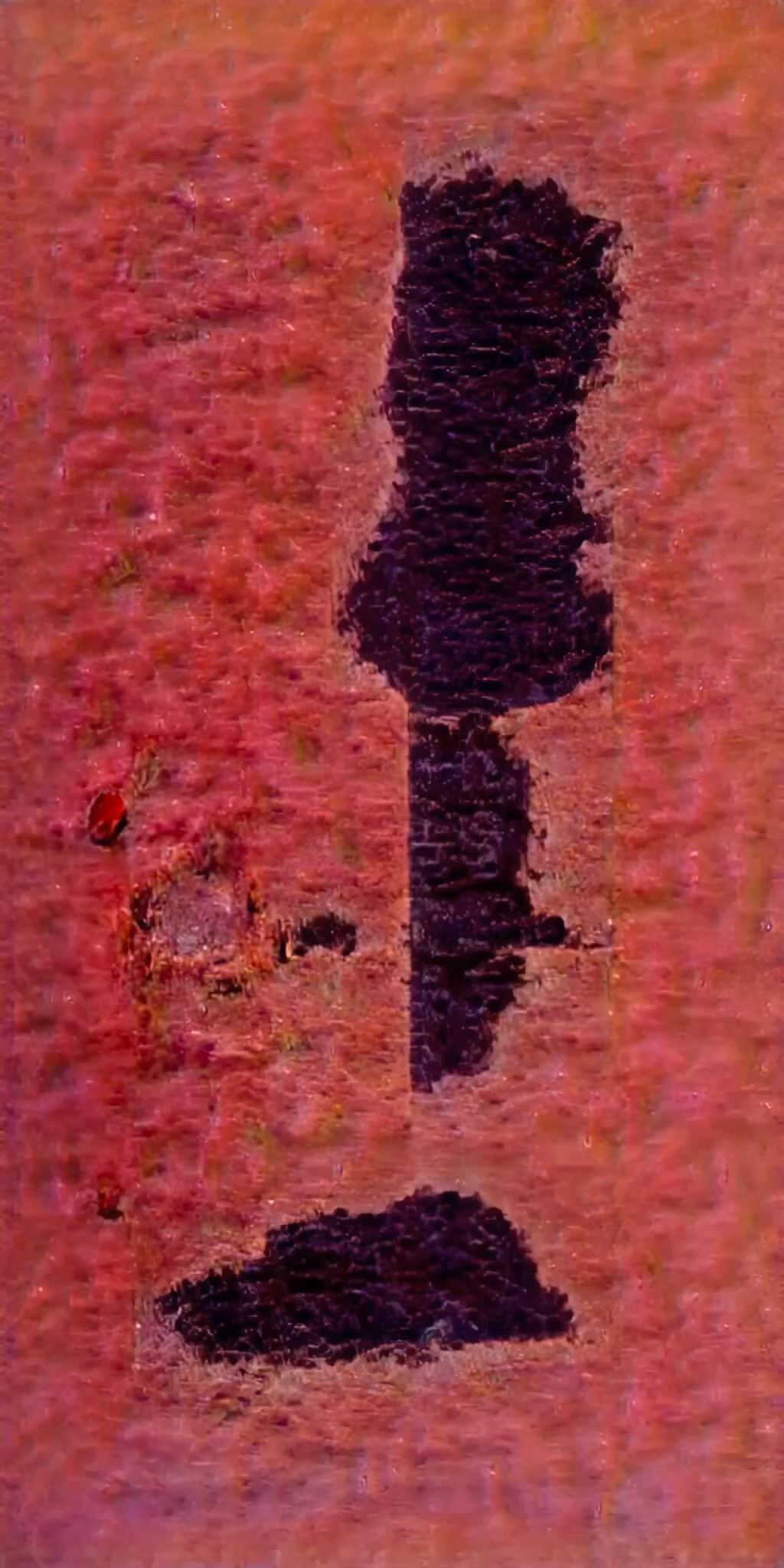}
    \caption{A sample frame produced by the proposed LAV (Latent Audio-Visual) system. \copyright StyleGAN2 model trained by Michael Friesen \cite{friesen}}
\end{figure}

\begin{figure*}[ht]
    \includegraphics[width=\textwidth]{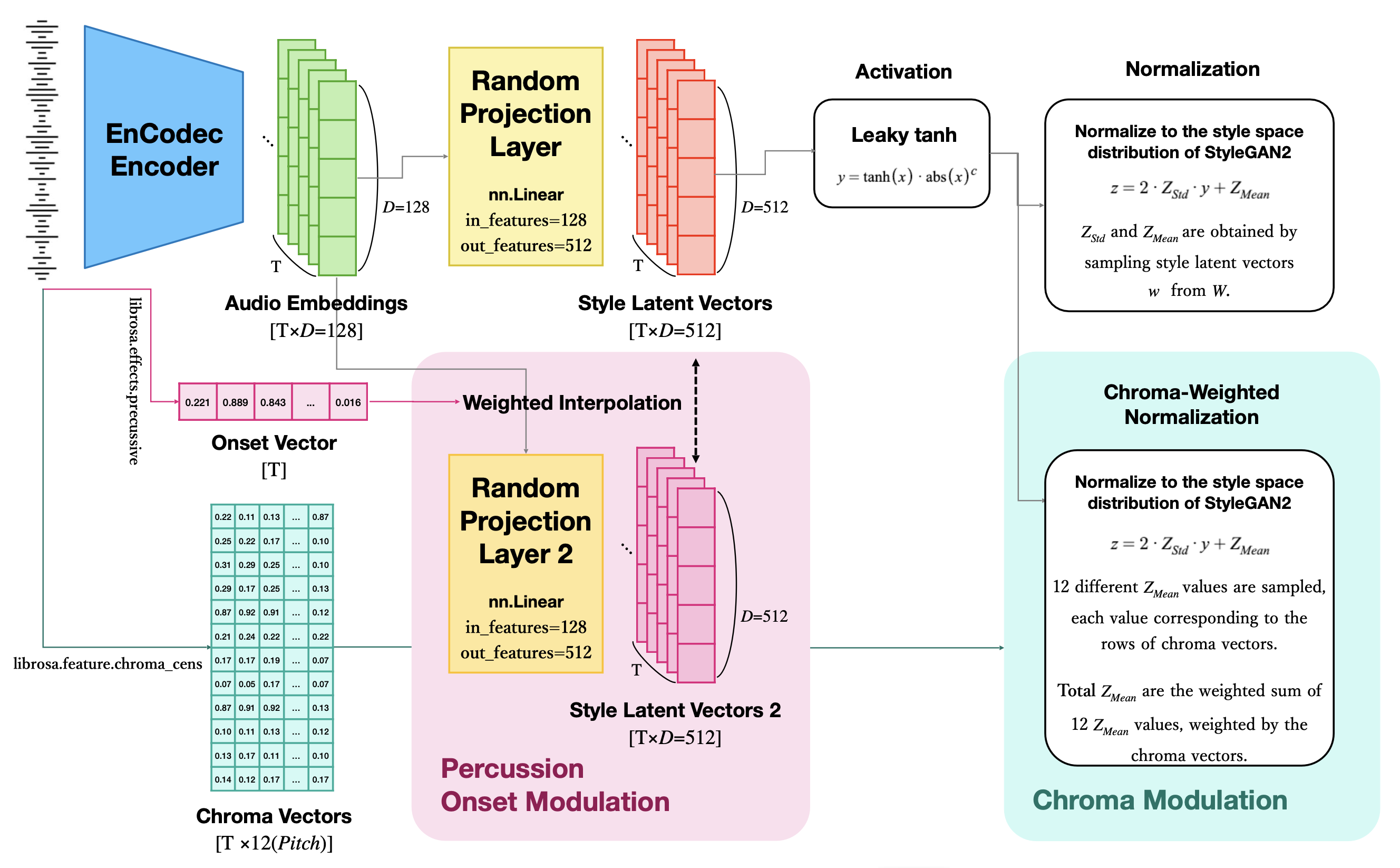}
    \caption{Overview of the proposed LAV (Latent Audio-Visual) pipeline.}
\end{figure*}

StyleGAN2 \cite{Karras2019stylegan2} uses a style-based generator architecture to enhance control and quality in image synthesis. At the core of its architecture is the style latent space $W$, an intermediate space derived from a mapping network that transforms input noise or latent vectors from the Gaussian input space $Z$. The style latent vector $w$ is a specific vector in this space that modulates the weights of convolutional layers in the generator through adaptive instance normalization (AdaIN). This modulation enables precise control over various image features, from coarse attributes like pose to finer details like texture, making $W$ crucial for generating high-quality, semantically consistent outputs.

Recent advancements in audio-driven visual synthesis have begun addressing this challenge, particularly with the application of StyleGAN \cite{DBLP:journals/corr/abs-1812-04948} for generative image synthesis. Notable contributions include \textit{Audio-reactive Latent Interpolations with StyleGAN} \cite{brouwer}, which modulates StyleGAN’s latent space using explicitly extracted audio features such as chroma and rhythmic onsets. Although effective in generating visually dynamic output, this approach requires the integration of hand-crafted audio features, which may limit semantic richness.

\textit{TräumerAI: Dreaming Music with StyleGAN} \cite{traeumer} takes a different approach by training a transformation layer to map audio features directly into StyleGAN2’s style latent space. This system pairs music with corresponding visual styles, focusing on creating seamless transitions across musical sections. However, this method relies on task-specific training, which introduces additional complexity and dependency on annotated data for effective performance.

In contrast, LAV (Latent Audio-Visual) eliminates the need for additional training layers by introducing a randomly initialized projection layer. This projection layer transforms audio embeddings directly into StyleGAN2’s style latent space via linear mapping. As proposed in \textit{A Foundation Model for Music Informatics} \cite{minz}, this approach not only simplifies the pipeline by removing the train process, but also ensures the preservation of high-level semantic information. By leveraging audio embeddings retrieved with pre-trained EnCodec encoder, a neural audio compression model which has the latent embedding sample rate of 50Hz unlike other foundation models, the transformed embeddings inherently retain semantic richness and can fluently and swiftly respond to the immediacy of music, effectively capturing and representing its essence. inherently retain semantic richness. This method overcomes the limitations of previous studies while expanding the possibilities for audio-visual synthesis.

\begin{figure*}
    \includegraphics[width=\textwidth]{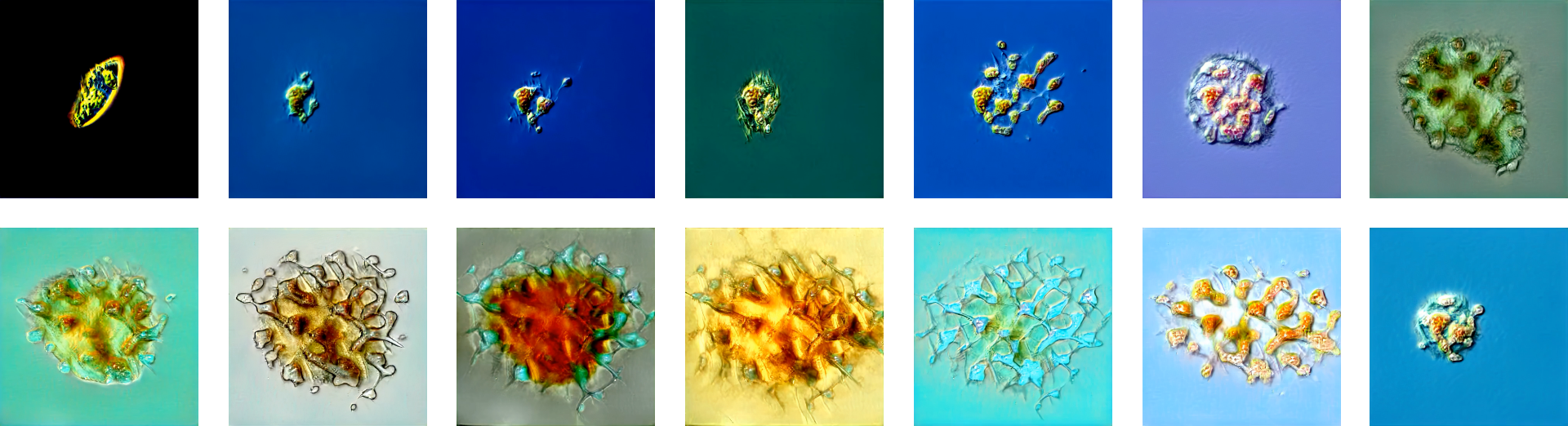}
    \caption{A sequence of frames produced by the proposed LAV (Latent Audio-Visual) system. \copyright StyleGAN2 model trained by Michael Friesen \cite{friesen}}
\end{figure*}

\section{Methodology}
 
\subsection{EnCodec Embedding Utilization}

EnCodec compresses audio into latent embeddings at a sample rate of 50Hz, with a latent dimension of $D=128$. Like other robust deep-learning based audio encoders, EnCodec embeddings inherently capture semantic audio features, such as timbre and structure, without requiring explicit extraction of chroma or rhythm.

\subsection{Latent Space Mapping}

A randomly initialized single projection layer maps EnCodec embeddings to StyleGAN2’s style latent space $W$ of dimensions $D=512$. As proposed in TräumerAI \cite{traeumer}, the transformed embeddings are normalized to the distribution of $W$ using $Z_{Std}$ and $Z_{Mean}$, which are obtained by sampling style latent vectors $w$ from $W$, with a controllable coefficient $y$ allowing for further adjustment of the normalization process. This transformation leverages the semantic integrity of EnCodec embeddings, with a leaky tanh activation function with controllable coefficient $c$ that preserves outliers and ensures compatibility with StyleGAN2’s distribution.

\subsection{Modulation by Musical Features}

As proposed in \textit{Audio-reactive Latent Interpolations with StyleGAN} \cite{brouwer}, modulating with hand-crafted features adds even more dynamic to the output. Musical features such as chroma and percussive onsets are extracted from the input audio using \textit{librosa} \cite{mcfee2015librosa}. 

\begin{itemize}
\item Onset Modulation: Percussive onset vectors dynamically weight outputs from two projection layers, capturing rhythmic patterns.
\item Chroma Modulation: Chroma vectors modulate the style latent space through weighted sums of 12 pitch-specific $Z_{Mean}$ values, enabling harmonic content to influence visual aesthetics.
\end{itemize}

\subsection{Hierarchical Style Smoothing}

To ensure smooth transitions in the generated images and prevent abrupt changes, an averaging window is applied to the style latent vectors used in the synthesis process. As proposed in \textit{TräumerAI} \cite{traeumer}, StyleGAN2’s synthesis network $g$ progressively refines image details as convolutional layers are applied, transitioning from coarse to fine-grained features. At each convolutional layer, the AdaIN module applies a corresponding style vector derived from the latent vector $w$. These convolutional layers are grouped into coarse, middle, and fine hierarchies, and the averaging window size is tailored to each group.

\section{Results and Demonstration}
The LAV system successfully transforms audio inputs into visually dynamic and semantically coherent outputs. Using pre-trained EnCodec embeddings, the system demonstrates remarkable capabilities in audio-visual synthesis:
\subsection{Visual Quality}
\begin{itemize}
\item High-resolution image generation that maintains visual fidelity while reflecting audio characteristics
\item Smooth transitions between frames through hierarchical style smoothing
\item Stable and consistent image generation across varied inputs
\end{itemize}
\subsection{Audio-Visual Mapping}
\begin{itemize}
\item Seamless translation of diverse audio inputs into semantically rich visual outputs
\item Comprehensive capture of audio semantics through EnCodec embeddings
\item Consistent performance across various musical genres and styles
\end{itemize}

\begin{figure*}
    \includegraphics[width=\textwidth]{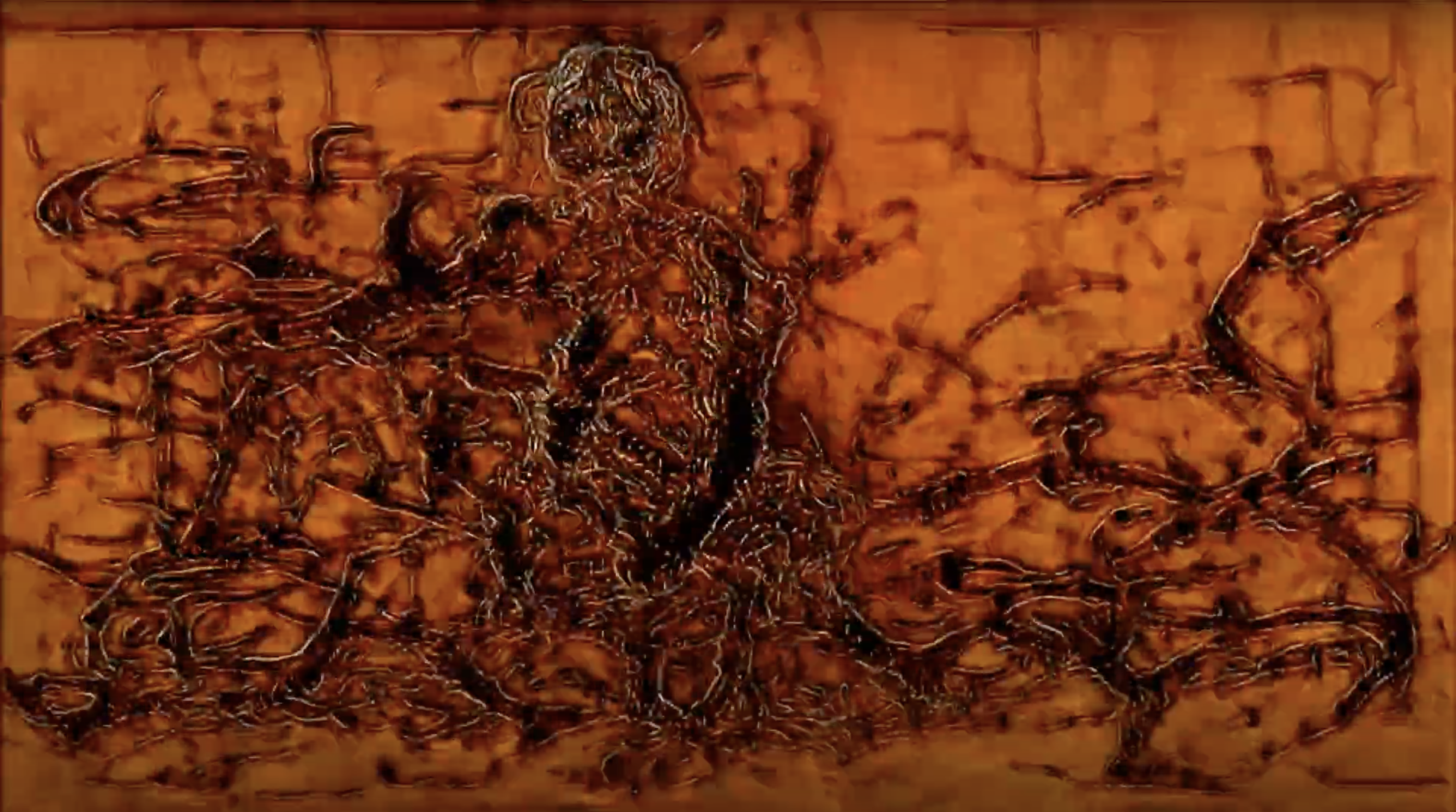}
    \caption{A sample frame produced by the proposed LAV (Latent Audio-Visual) system. \copyright StyleGAN2 model trained by Krrrl \cite{krrrl}}
\end{figure*}

\section{Conclusion}
This work introduces LAV (Latent Audio-Visual), a novel framework that bridges neural audio compression and generative visual synthesis. By transforming EnCodec's semantically rich embeddings into StyleGAN2's style latent space through a randomly initialized linear mapping, the system achieves high-quality audio-visual translation without additional training requirements. The framework demonstrates the potential of using pre-trained audio compression models for artistic and computational applications, establishing a foundation for future developments in audio-visual synthesis, particularly in areas where semantic coherence between audio and visual elements is paramount.

\section{Acknowledgement}
This work was supported by the Ministry of Education of the Republic of Korea and the National Research Foundation of Korea (NRF-2024S1A5C3A03046168).

\bibliographystyle{isea}
\bibliography{isea}

\section{Author Biography}
Jongmin Jung is an AI researcher and creative technologist pursuing a master's degree in Artificial Intelligence at Sogang University. Having earned a Bachelor of Art \& Science (BAS) in Art \& Technology, he works at the intersection of AI and creative expression, with a focus on generative audiovisual art, music generation models, automatic music transcription (AMT), optical music recognition (OMR) and score-image to audio generation task.
He actively collaborates with diverse artists, including musicians, tattooists, and visual creators, to integrate cutting-edge AI technology into artistic practices. Through his development of custom audiovisual pipelines and sound generation projects, Jongmin explores novel ways to merge machine learning with artistic innovation.

\end{document}